\documentclass[reprint,amsmath,amssymb,aps,floatfix]{revtex4-2}
\usepackage{graphicx}
\usepackage{dcolumn}
\usepackage{bm}
\usepackage{bbm}
\usepackage{float}
\usepackage{makecell}

\usepackage{cleveref}
\usepackage{xr}
\usepackage{soul}

\usepackage{color,soul}
\newcommand{\rev}[1]{\textcolor{black}{#1}}

\makeatletter
\newcommand*{\addFileDependency}[1]{
  \typeout{(#1)}
  \@addtofilelist{#1}
  \IfFileExists{#1}{}{\typeout{No file #1.}}
}
\makeatother
\newcommand*{\myexternaldocument}[1]{
    \externaldocument{#1}
    \addFileDependency{#1.tex}
    \addFileDependency{#1.aux}
}
\myexternaldocument{supplementary}

\newcommand{\chieff}{\ensuremath{\chi_{\text{FHP}}(T)\,}}
\newcommand{\Rzero}{\ensuremath{\mathcal{R}_0\,}}
\newcommand{\Rone}{\ensuremath{\mathcal{R}_1'\,}}
\newcommand{\Rtwo}{\ensuremath{\mathcal{R}_2\,}}
\newcommand{\Rthree}{\ensuremath{\mathcal{R}_3'\,}}

\begin{document}

\preprint{}

\title{\rev{Asymmetry in Polymer-Solvent Interactions Yields Complex Thermoresponsive Behavior}}

\author{Satyen Dhamankar}
\author{Michael A. Webb}%
 \email{mawebb@princeton.edu}
\affiliation{%
Department of Chemical and Biological Engineering, Princeton University, Princeton, NJ 08544}%

\date{\today}

\begin{abstract}
Thermoresponsive polymers hold both fundamental and technological importance, but the essential physics driving their intriguing behavior is not wholly understood.
We introduce a lattice framework that incorporates elements of Flory-Huggins solution theory and the $q$-state Potts model to study the phase behavior of polymer solutions and single-chain conformational characteristics.
Importantly, the framework does not employ any temperature- or composition-dependent parameters.
With this minimal Flory-Huggins-Potts framework, we show that orientation-dependent interactions, specifically between monomer segments and solvent particles, are alone sufficient to observe upper critical solution temperatures, miscibility loops, and hourglass-shaped spinodal curves.
Signatures of emergent phase behavior are found in single-chain Monte Carlo simulations, which display heating- and cooling-induced coil-globule transitions linked to energy fluctuations.
\rev{The model also capably describes a range of experimental systems. 
This work provides new insights regarding the microscopic physics that underpin complex thermoresponsive behavior in polymers.}

\end{abstract}

\maketitle


Thermoresponsive polymers (TRPs) drastically alter structure, functionality, and/or stability upon changes in temperature.\cite{R:2016_Karimi_Temperature, R:2017_Kim_Thermo, R:2011_Ward_Thermoresponsive} 
They are appealing candidates for many applications, such as drug delivery, sensing, and purification.\cite{R:2009_Davis_Force, R:2012_Colson_Biologically,R:1982_Tanaka_Collapse, R:2013_Thevenot_Magnetic, R:2013_Zhai_Stimuli, R:2019_Mukherji_Soft, R:2010_Stuart_Emerging, R:2020_Mukherji_Smart,R:2022_Xu_Thermoresponsive} 
Their physics also have implications for biological systems, with parallels to cold and warm denaturation of proteins, the hydrophobic effect, the formation of biomolecular condensates, and folding of DNA/RNA nanostructures.\cite{R:2020_Zeng_Connecting, R:1967_Scarpa_Slow,R:1989_Shakhnovich_Theory, R:2000_Rios_Putting, R:2000_Gelbart_DNA,R:2017_Shin_Liquid,R:2019_Dignon_Temperature,R:2014_Geary_single,R:2017_Han_Single,R:2021_Jia_DNA} 
Manipulating how polymer-based materials respond to temperature could also enhance control over supramolecular assemblies and improve particle engineering.\cite{R:2010_Yoo_Polymer,R:2015_Lewandowski_Dynamically,R:2022_Bupathy_Temperature}
Consequently, understanding and modeling TRPs is of significant interest. 

No self-contained theoretical framework currently captures the range of  behaviors observed in TRPs.
Flory-Huggins solution theory\cite{R:1949_Flory_Configuration,R:1953_Flory_Principles} (FH) is a simple conceptual starting point, but balancing short-ranged interactions against chain and solvent entropy only yields upper critical solution temperatures (UCST) and heating-induced globule-coil transitions (GCT).\cite{R:1953_Flory_Principles,R:1949_Flory_Configuration,R:1979_Gennes_Scaling,R:1975_Gennes_Collapse}
Emergence of lower critical solution temperatures (LCST) or heating-induced coil-globule transitions (CGT) has been addressed by theoretical extensions,\cite{R:1976_Lacombe_Statistical,R:1976_Sanchez_elementary,R:1978_Sanchez_Statistical,R:1987_Panayiotou_Lattice, R:1991_Panayiotou_Hydrogen,R:1990_Veytsman_Are,R:2008_Simmons_Model,R:2010_Simmons_Pressure, R:2021_Dahanayake_Hydrogen, R:1990_Matsuyama_Theory,R:1997_Bekiranov_Solution, R:2019_Choi_LASSI:,R:2008_Oh_Closed,R:1991_Hu_Double, R:2012_Clark_LCST,R:1978_Andersen_Theory} 
\rev{which either introduce adjustable parameters or impose \emph{ad hoc} temperature-dependence to existing ones. 
Some representative strategies include utilizing void sites and instituting surface-area effect corrections\cite{R:1976_Sanchez_elementary}, incorporating secondary lattices with temperature-dependent perturbative interactions\cite{R:2008_Oh_Closed, R:2012_Clark_LCST}, invoking specific associative interactions,\cite{R:2015_Dudowicz_Communication:} or prescribing  an equation of state.\cite{R:1991_Panayiotou_Hydrogen,R:1978_Andersen_Theory,R:2012_Clark_LCST} 
Although these approaches are powerful and expressive, they often lack a firm microscopic basis, which inhibits validation, extension, interpretation, and application.
By contrast, 
molecular simulation analysis has pointed to molecular size effects, hydrogen-bonding, and other orientation-dependent interactions as important for thermoresponsive behavior.\cite{R:1996_Lee_Two,R:2000_Rios_Putting, R:1996_Jeppesen_Single,R:2020_Martin_Valence,R:1997_Raos_Macromolecular,R:1998_Panagiotopoulos_Phase,R:2015_Zhang_Polymers,R:2021_Arsiccio_Protein,R:2015_Abbott_temperature,R:2021_Dhamankar_Chemically}
However, the chemical specificity obfuscates general comprehension and presents little opportunity to isolate the minimal and necessary set of interactions. 
Thus, 
questions remain regarding the microscopic physics of TRPs.}

In this Letter, we introduce a minimal framework that combines elements of Flory-Huggins theory\cite{R:1949_Flory_Configuration,R:1953_Flory_Principles} and the $q$-states Potts model\cite{R:1952_Potts_Some, R:1980_Walker_Theory} to understand phase behavior and the CGT in TRPs.
\rev{A key feature of this Flory-Huggins-Potts (FHP) framework is that all model parameters are linked to a Hamiltonian that transparently depends on pairwise interaction energies amongst monomer and solvent particles.
By consequence, derived models are free of temperature- and composition-dependent parameters, permitting both analytical analysis and commensurate investigation with molecular simulation.
Mean-field analysis reveals that  asymmetry in monomer-solvent interactions is alone sufficient to produce intricate phase behavior, such as miscibility loops and hourglass-shaped phase envelopes, which are not captured by native FH. 
Furthermore, Monte Carlo simulations establish a connection between anticipated phase behavior and single-chain conformational characteristics, which enables microscopic analysis that extends our mean-field understanding.
The FHP framework is also shown to reproduce diverse phase-coexistence data for several experimental systems.}
These findings not only highlight energetic asymmetries as important physics underlying TRPs but also establish a self-contained, conceptual framework that may be broadly applied to TRPs or extended to other stimuli-responsive systems. 

\begin{figure}[htbp!]
\includegraphics{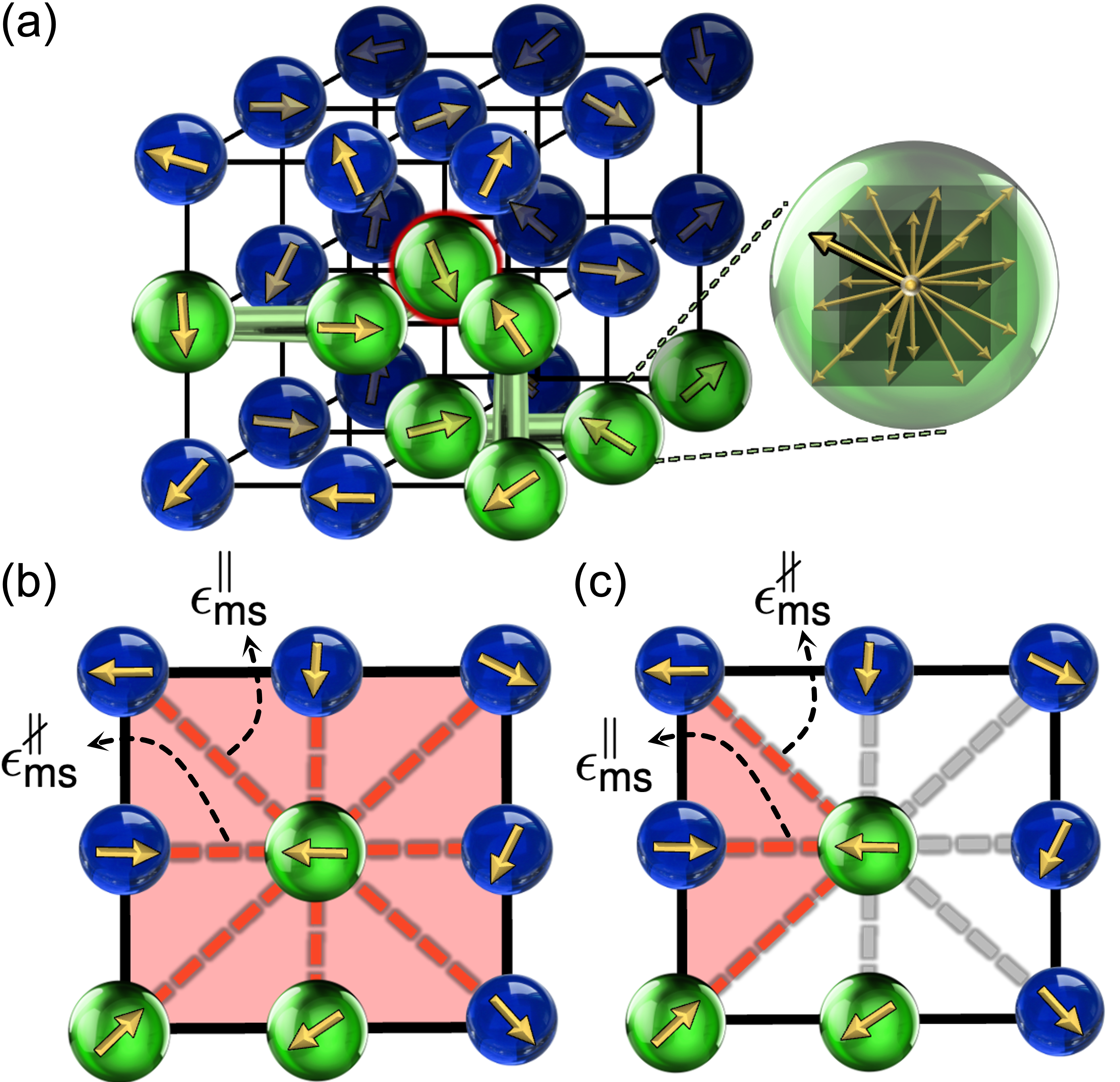}
\caption{
Schematic of the Flory-Huggins-Potts framework. 
(a) The local neighborhood of a site (red outline) for a simple cubic lattice. Monomer (m) segments and bonds are in green, and solvent particles (s) are blue.
(b, c) Two-dimensional illustrations of possible manifestations of orientation-dependent interactions. In (b), aligned interactions are permitted with all neighboring particles, which may capture physics akin to hydrogen-bond networks. In (c), aligned interactions are permitted with only a subset of neighboring particles, which may reflect specific and exclusive associative interactions. In both panels, particles adjoined by red lines in the shaded region can engage in aligned ($\parallel$) interactions; those adjoined by gray only exhibit misaligned ($\nparallel$) interactions.}\label{fig:Fig1_LatticeViz}
\end{figure}

We consider a lattice completely occupied by either solvent particles or  monomer segments (Fig. \ref{fig:Fig1_LatticeViz}a). 
Like FH, short-ranged, pairwise interactions exist between particles, and polymers consist of bonded monomer segments.
Like a $q$-state Potts model, particles possess orientations that can influence pairwise energies.
The system energy is generally
\small
\begin{align}
    \mathcal{H} &= \frac{1}{2} \sum _{i=1} ^N \sum _{j \in \mathcal{N}(i)} \varepsilon \left(\alpha _i, \alpha _j, \hat{\sigma}_i, \hat{\sigma}_j \right) + \sum_{k=1}^m \sum_{l=1}^{M-1} V\left(\vec{r}_{l+1}^{(k)},\vec{r}_{l}^{(k)}\right) \label{eq:genenergy}
\end{align}
\normalsize
with
\begin{align}\label{eq:bond}
    V\left(\vec{r}_{l+1}^{(k)},\vec{r}_{l}^{(k)}\right)  = 
    \begin{cases}
        0 & \text{if } \left|\vec{r}_{l+1}^{(k)}-\vec{r}_{l}^{(k)}\right| \in \mathcal{N}(\mathcal{O})  \\
        \infty & \text{otherwise}
    \end{cases} 
\end{align}
where $\alpha_i$ denotes the species (monomer or solvent) and $\hat{\sigma}_i$ denotes the unit-vector orientation of the the particle occupying the $i$th lattice site; $\varepsilon(\cdot)$ is a pairwise energy function of such variables; $N$ is the total number of lattice sites; $m$ is the number of polymer chains; $M$ is the number of monomer segments per chain; $\vec{r}_{l}^{(k)}$ is the position of the $l$th monomer segment in the $k$th polymer chain; and $V(\cdot)$ is a potential function that ensures bonded monomer segments occupy neighboring positions on the lattice (their distance is in the neighborhood of the origin, $\mathcal{N}(\mathcal{O})$).
Eq. {\eqref{eq:genenergy}} is distinguished from FH by the influence of particle orientations on pairwise energies.

To simplify, we decompose 
\begin{equation}
\varepsilon(\alpha _i, \alpha _j, \hat{\sigma}_i, \hat{\sigma}_j) = \epsilon _{\alpha _i \alpha _j} ^{\nparallel} + \Lambda(i,j) \Delta_{\alpha _i \alpha _j} 
\end{equation}
where $\Delta_{\alpha _i \alpha _j} = \epsilon _{\alpha _i \alpha _j} ^{\parallel} - \epsilon_{\alpha _i \alpha _j}^\nparallel$ is the difference between interaction energies for species $\alpha _i$ and $\alpha _j$ in an ``aligned'' ($\parallel$) versus  ``misaligned'' ($\nparallel$) state, and $\Lambda(i,j)\in[0,1]$ functionally determines the extent to which particles at lattice sites $i$ and $j$ are aligned.
By convention and without loss of generality, we presume $\Delta_{\alpha _i \alpha _j}\leq 0$ such that aligned interactions are energetically favorable than misaligned. 
We envision that interaction asymmetries (i.e., differences in energy between aligned and misaligned states) arise from detailed chemical structure and molecular geometry. Two contrasting scenarios are depicted in  Fig. \ref{fig:Fig1_LatticeViz}b, which is inspired by  hydrogen-bonding networks\cite{R:2012_Deshmukh_Role}, and Fig. \ref{fig:Fig1_LatticeViz}c, which is inspired by specific associative interactions\cite{R:2012_Martinez_Rethinking,R:2023_Danielsen_Phase} (e.g., pi-pi stacking, metal–ligand, electrostatic).
The primary distinction between these is the fraction of neighboring particles capable of aligned interactions, denoted $p_v$ (red shading).

To gain insight into phase behavior, we consider the mean-field free energy. \rev{Following principles of FH and Boltzmann statistics (Supplemental Information, Section S1),  
the free energy of mixing per particle is
\begin{align}\label{eq:Fmix}
    \frac{\Delta\bar{F}_ {\text{mix}}}{k_\text{B} T} = \chieff \varphi(1-\varphi) +  \frac{\varphi}{Mv_m} \ln \varphi + \frac{(1-\varphi)}{v_s}\ln (1-\varphi)
\end{align}
where $\varphi$ is the volume fraction of the polymer, $M$ is the degree of polymerization, $v_m$ is the molar volume of a monomer segment, and $v_s$ is the same for a solvent molecule (or group thereof).}
Furthermore,  
\begin{equation}\label{eq:chieff}
    \chi _{\text{FHP}} (T) \equiv  \chi _{\text{FH}}(T) + \Tilde{\chi} (T)
\end{equation}
where
\begin{equation}\label{eq:chiFH}
    \chi _{\text{FH}}(T) \equiv \frac{(z-2)}{k_\text{B} T} \left( \epsilon _{\text{ms}} ^{\nparallel} - \frac{1}{2}\left(\epsilon _{\text{mm}} ^{\nparallel} + \epsilon _{\text{ss}} ^{\nparallel} \right) \right)
\end{equation}
is essentially equivalent to a FH interaction parameter for a lattice with coordination number $z$ and pairwise energies set at the misaligned energy scale, and
\begin{equation}\label{eq:chipotts}
    \Tilde{\chi} (T) \equiv \frac{(z-2)}{k_\text{B} T} p_v \left( \Tilde{\Delta}_{\text{ms}}(T) - \frac{1}{2}\left(\Tilde{\Delta}_{\text{mm}}(T) + \Tilde{\Delta}_{\text{ss}}(T) \right) \right)  
\end{equation}
accounts for asymmetry between aligned and misaligned interactions and their relative prevalence.
In particular, the $\Tilde{\Delta}_{ij}$ terms amount to per-site free-energy corrections between aligned and misaligned states given by
\begin{equation}\label{eq:correction}
 \Tilde{\Delta}_{ij} (T) = \frac{\Delta _{ij}}{1+ r\exp(\Delta _{ij}/k_{\text{B}}T)}
\end{equation}
where $r=\frac{1-p_{\Omega}}{p_{\Omega}}$ is the fraction of possible misaligned to aligned microstates for a given pair of particles.

\rev{At this stage, we remark that FHP may appear deceptively complex.
Eq. {\ref{eq:Fmix}} itself is functionally analogous to the FH free energy of mixing, yet Eqs. {\eqref{eq:chieff}}-{\eqref{eq:correction}} seemingly introduce five parameters beyond FH ($\Delta_\text{ms}$, $\Delta_\text{ss}$, $\Delta_\text{mm}$, $r$, and $p_v$) belying simplicity.
However, neither $p_v$ nor $r$ are ``free'' parameters as they directly relate to the physical interactions (i.e., what constitutes aligned vs. misaligned) and follow from a defined Hamiltonian.
This potentially leaves twice as many energy parameters compared to original FH. 
However, we will show that a minimal model to observe complex phase behavior only requires $\Delta_\text{ms}<0$ (with $p_v$ and $r$ as nonzero).}

\begin{figure}[!htbp]
\includegraphics{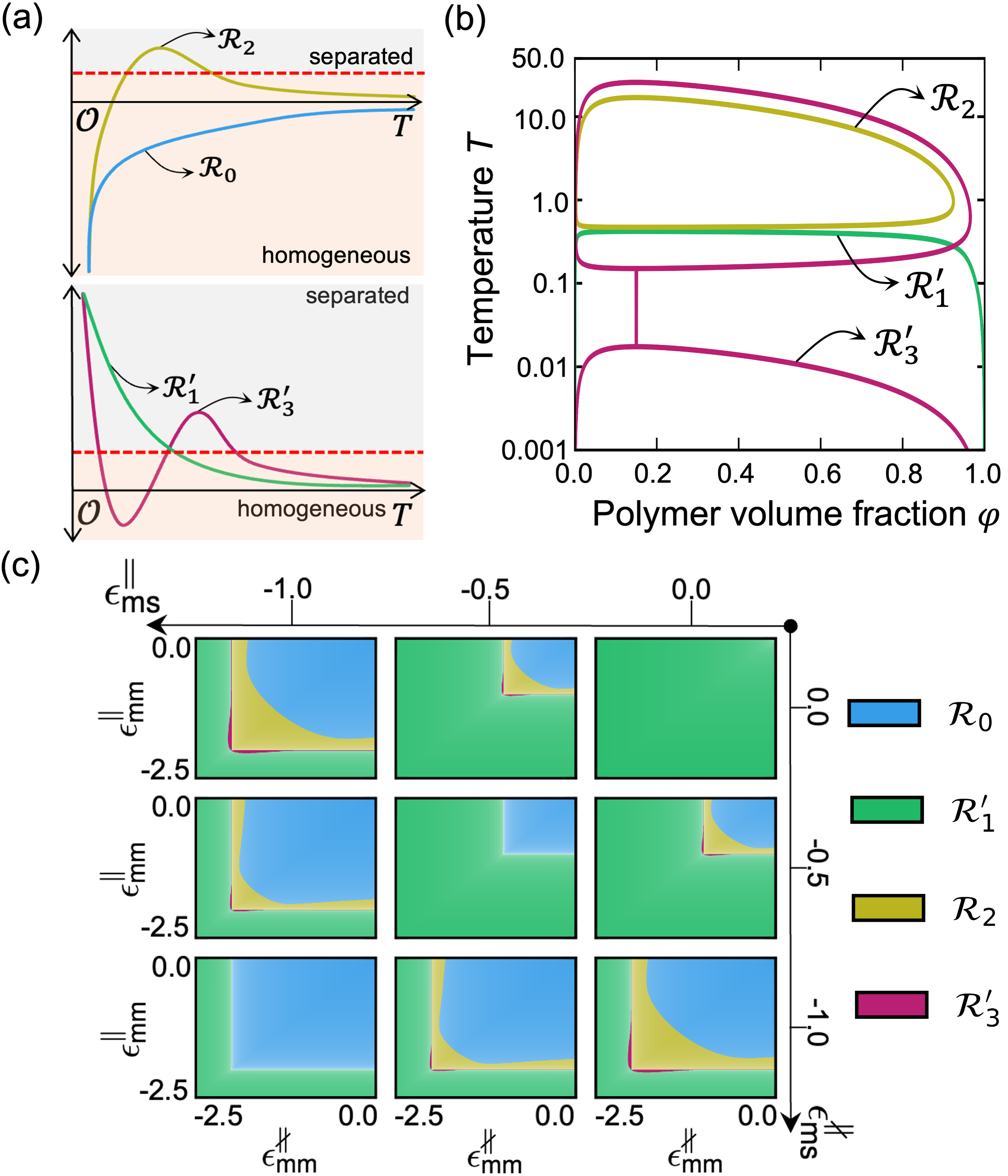}
\caption{Phase behavior of polymer solutions described by the Flory-Huggins-Potts framework. 
Representative temperature-volume fraction ($T-\varphi$)  for a polymer with degree of polymerization $M=32$, monomer molar volume $v_m = 1$ and solvent molar volume $v_s = 1$.
(a) Qualitative examples of \chieff for classification of phase behavior. Systems from \Rzero are expected to be homogeneous, while those from \Rone, \Rtwo, and {\Rthree} respectively feature upper critical solution temperatures, miscibility loops, and hourglass-shaped phase envelopes.
(b) Spinodal curves on temperature-volume fraction ($T-\varphi$) diagrams for models with parameters that yield \Rone, \Rtwo, and {\Rthree}; parameters for {\Rzero} yield only homogeneous phases.
(c) Classification of phase behavior as a function of $\epsilon _{\text{mm}}^{\parallel}$ and $\epsilon _{\text{mm}}^{\nparallel}$ (inner plots) at various values of $\epsilon _{\text{ms}}^{\parallel}$ and $\epsilon _{\text{ms}}^{\nparallel}$ (outer axes).
Results are obtained from mean-field analysis with $(\epsilon _{\text{ss}} ^{\parallel}, \epsilon _{\text{ss}} ^{\nparallel}, p_v, p_{\Omega})$ =(0, 0, 1, 0.25) and $z=26$. 
Additional parameter-sets are examined in Supplemental Information (Figs. S1, S2).} \label{fig:Fig2_PhaseBehavior}
\end {figure}

\rev{To illustrate, we characterize the phase behavior for a select set of models from four FHP-derived regimes.
In FHP, we anticipate four regimes of qualitatively distinct \chieff behavior -- denoted as {\Rzero}, {\Rone}, {\Rtwo}, or {\Rthree} (Fig. {\ref{fig:Fig2_PhaseBehavior}a}) -- that determine observed phases.}
The notation is such that the subscript indicates the number of homogeneous-to-phase-separated transitions, a prime ($'$) indicates a phase-separated state at low $T$, and no prime indicates a homogeneous state at low $T$. 
The spinodals in 
Fig. {\ref{fig:Fig2_PhaseBehavior}b} illustrate how the FHP framework exhibits  complex phase behavior relative to FH for specific parameter-sets  representative of {\Rone}, {\Rtwo}, and {\Rthree}.
\rev{While {\Rone}  displays UCST, which is well-captured in FH, {\Rtwo} and {\Rthree} respectively display a miscibility loop and an hourglass-like phase envelope, which is usually recovered by invoking an equation-of-state or otherwise presuming temperature-dependent parameters.
Additionally, these models use $\Delta_\text{mm} = \Delta_\text{ss} =0 $, which suggests that the introducing interaction asymmetry to monomer-solvent interactions ($\Delta_\text{ms} <0$) is alone sufficient to obtain miscibility loops and hourglass-like phase envelopes.}

\rev{Conversely, only introducing asymmetry to monomer-monomer interactions (or likewise solvent-solvent) does not result in as rich phase behavior.}
This is conveyed in Fig. \ref{fig:Fig2_PhaseBehavior}c, which provides a ``hyperphase diagram'' summarizing $\chieff$  behavior across a much broader parameter-space rather as opposed to the select cases in Fig. \ref{fig:Fig2_PhaseBehavior}b.
While the preponderance of parameter-space yields {\Rzero} (blue) or {\Rone} (green) behavior, {\Rtwo} (miscibility loops, yellow) and {\Rthree} (hourglass envelopes, magenta) emerge within intermediate energetic regimes only if misaligned and aligned monomer-solvent interactions are unequal. 
Overall, these results show how the balance of both homotypic and heterotypic interactions as well as misaligned and aligned interactions give rise to nuanced phase behavior, but asymmetric interactions between monomer and solvent are essential to observe multiple phase transitions within the FHP framework. 

\begin{figure}[t]
\includegraphics{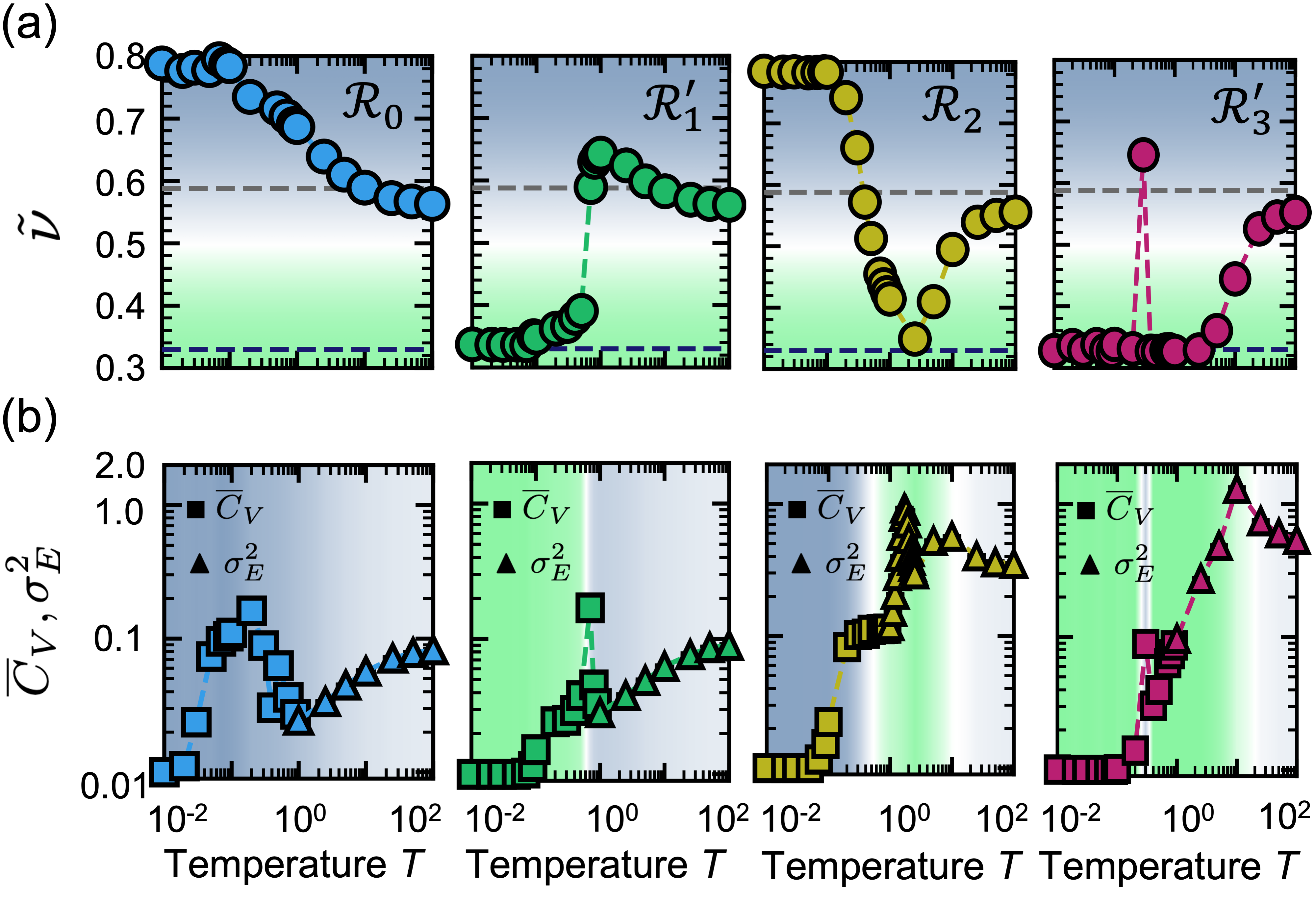}
\caption{Analysis of single-chain Monte Carlo simulations. 
(a,b) Temperature dependence of (a) intra-chain scaling exponent $\Tilde{\nu}$ and (b) energy fluctuations using parameters associated with {\Rzero}, {\Rone}, {\Rtwo} and {\Rthree}.
In (a), the horizontal dashed lines are guides to the eye for globular scaling (black) and excluded-volume statistics (gray).
In (b), the data transitions such that heat capacity $\overline{C}_V$ is shown for $T<1.0$ and fluctuations in total energy $\sigma_E^2$ is shown for $T\geq 1.0$; the data are shifted by $0.01$ to allow for logarithmic scaling.
The background color is based on $\tilde{\nu}$ to emphasize conformations that are expanded (blue) or globular (green) relative to the ideal chain (white).
 Error bars are smaller than the symbol size.
}\label{Fig5:FluctuationsInEnergy}
\end{figure}

\rev{Because the parameters within FHP are traceable to a well-defined Hamiltonian, molecular simulation can be used to ascertain any connection between single-chain structure and solution thermodynamics (as established for FH\cite{R:1998_Panagiotopoulos_Phase,R:2014_Wang_Theory}).}
Monte Carlo (MC) simulations  are used to characterize the temperature-dependent behavior of single polymer chains using parameters from each of \Rzero, \Rone, \Rtwo, and \Rthree.
The simulations use $M=32$ and a $\Lambda$ inspired by hydrogen-bonding networks
\begin{equation}\label{eq:LambdaCorr}
\Lambda_{\text{corr}}(i,j)  = \Theta (\hat{\sigma} _i \cdot \hat{\sigma} _j - \delta), 
\end{equation}
which permits $p_v=1$ and for $\delta = \cos 1$ gives $r = 3$ (see also Fig. {\ref{fig:Fig1_LatticeViz}b}).
To probe single-chain conformations, we extract intra-chain scaling exponents $\tilde{\nu}$ from
\begin{align}
    \left\langle |\vec{r}_i - \vec{r}_j| ^2 \right\rangle \propto |i-j| ^{2\tilde{\nu}}
\end{align}
and compare with known scaling laws (e.g., $R_\text{g} \propto M^{1/2}$ for ideal chains, $R_\text{g} \propto M^{1/3}$ for globules). 


Fig. \ref{Fig5:FluctuationsInEnergy}a shows that distinct single-chain conformational behavior is observed when parameters are taken from different phase behavior regimes.  
At low $T$, systems from  {\Rzero} and {\Rtwo}  exhibit scaling indicative of good-solvent conditions ($\tilde{\nu} \geq 2/3$), while systems from {\Rone} and {\Rthree} display more globular scaling; 
this correlates with {\Rzero} and {\Rtwo} having a single homogeneous phase and {\Rone} and {\Rthree} featuring phase-separated states.
Behavior between {\Rzero} and {\Rtwo} differs upon heating. 
While $\tilde{\nu}$ from {\Rzero} gradually approaches excluded-volume statistics ($\tilde{\nu} \sim 0.588$), $\tilde{\nu}$ from {\Rtwo} collapses to a globule before expanding in the high-$T$ limit, which aligns with the physics of a miscibility loop. 
Similar explanations apply for {\Rone} and {\Rthree}. The singular, sharp globule-coil transition for $\tilde{\nu}$ from {\Rone} with asymmetric monomer-monomer interactions is consistent with a UCST, but a sharp transition is only observed with sufficiently asymmetric interactions.
For {\Rthree}, $\tilde{\nu}$  begins as globular then crosses the ideal-chain limit three times--a progression consistent with an hourglass-like phase envelope.
In contrast, $\tilde{\nu}$ obtained with FH 
only tends gradually and monotonically to excluded-volume statistics upon heating.
Thus, anisotropic interactions can drive sharp, thermally induced conformational transitions -- reminiscent of the trends seen in phase behavior.

The observed conformational transitions are also accompanied by drastic shifts in underlying energy fluctuations (Fig. {\ref{Fig5:FluctuationsInEnergy}}b). 
For systems from {\Rone}, the single CGT is marked by maxima in the heat capacity $\overline{C}_V$ and energy fluctuations $\sigma^2_E$  within a narrow temperature range wherein $\tilde{\nu}\sim 0.5 $ (white region);
similar observations hold for the low-$T$ CGT in {\Rtwo} and {\Rthree}.
In {\Rtwo} and {\Rthree}, a second higher-$T$ transition from globular to excluded-volume statistics is captured by a maximum in $\sigma^2_E$.
These fluctuations notably accompany compositional and orientational changes within the polymer's local environment (Supplemental Information, Figs. S2 and S3) and are thus not reproducible by FH but are inherent to FHP.

\begin{figure}[!htbp]
\includegraphics[keepaspectratio,width=\columnwidth]{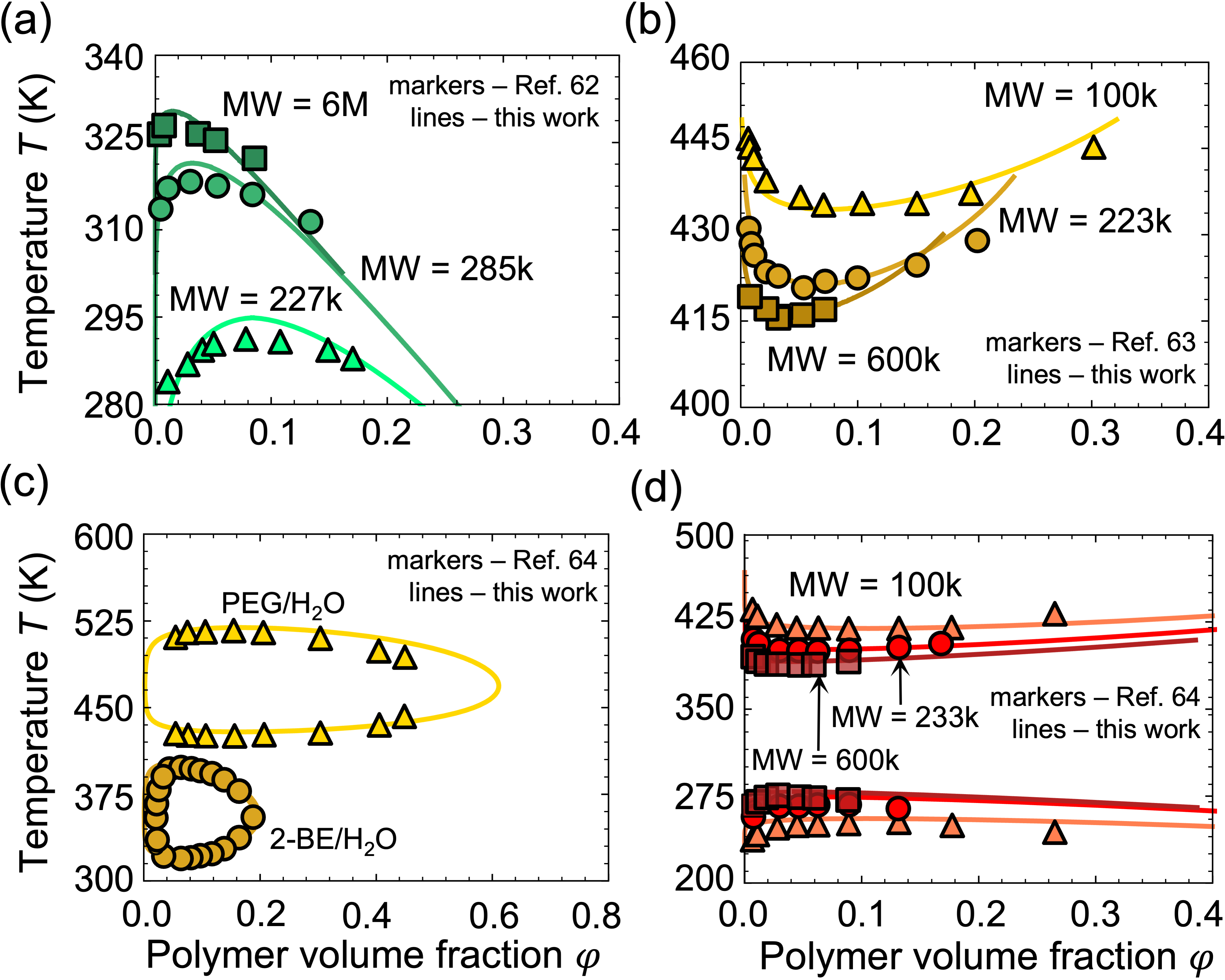}
\caption{\rev{Comparison to experimental coexistence data for diverse systems. Temperature-volume fraction ($T-\phi$) diagrams for systems that display (a) upper critical solution temperatures, such as polyisbutylene (PIB) in diisobutyl ketone at several molecular weights (MW); (b) lower critical solution temperatures, such as polystyrene in ethyl acetate at several MW; (c) miscibility loops, such as polyethylene glycol (PEG) (3350 amu) in water and 2-butoxyethanol (2-BE) in water; and (d) hourlgass-like phase behavior, such as polystyrene in tert-butyl acetate at several MW. 
The data in (a)-(d) are respectively sourced from Refs. \cite{R:1952_Shultz_Phase}, \cite{R:1993_Bae_Representation}, and \cite{R:1979_Arlt_Liquid} and were also compared to theory proposed in Ref. \cite{R:2008_Oh_Closed, R:2013_Oh_Molecular}. 
All molecular weights are reported in Daltons. All parameters are reported in the Supplemental Information (Table 4).}}
\end{figure}

\rev{Finally, the FHP framework can quantitatively capture the behavior of experimental systems. 
This is illustrated in Fig. 4, which 
compares FHP results with experimental data for various solutions that collectively showcase UCST, LCST, miscibility loops, and hourglass (UCST + LCST) phase envelopes.
These same systems were examined by Bae and Oh to evaluate a proposed double-lattice model.\cite{R:1993_Bae_Representation} 
Here, for each set of experimental data, the FHP parameters were optimized using the covariance matrix adaptation evolutionary strategy.\cite{R:2001_Hansen_Completely}
It is notable that FHP captures the experimental data well without any temperature-dependent parameters or externally imposed equation-of-state, which is distinct from similarly expressive  prior theoretical works.\cite{R:1993_Bae_Representation, R:1978_Sanchez_Statistical, R:2012_Clark_LCST}
Although the quality of fits here does not firmly exclude the importance of other key interactions, the capacity to represent some facets of experimental data is not only an important demonstration for FHP but bolsters the case that energetic asymmetries, which distinguish FHP, are important to the physics of real systems.}

In conclusion, we have introduced the Flory-Huggins-Potts (FHP) framework to describe and understand thermoresponsive polymers. 
Remarkably, this approach, which simply extends Flory-Huggins (FH) theory with orientation-dependent interactions, showcases diverse temperature-dependent phase behavior, including miscibility loops and hourglass-like phase envelopes. 
\rev{In this way, FHP can capture the experimental data of diverse systems.
This expressive capacity emerges without any \emph{ad hoc} functional dependencies on temperature; the effects are transparently attributable to microscopic theory of interactions. 
Here,
this enabled investigation of how polymer solution thermodynamics connect to single-chain conformational behavior.
We find overall strong correspondence between single-chain conformational behavior (from molecular simulation) and macroscopic phase transitions (deduced from mean-field analysis), with multiple GCT transitions observed for a parameters that yield hourglass-like phase envelopes, for example. 
A key result from this analysis is that asymmetry in heterotypic (e.g., monomer-solvent) interactions is sufficient to induce complex thermoresponsive behavior, such as heating-induced CGT.}

\rev{In the future, FHP can be extended and applied to other physical and chemical systems. 
Firstly, it can serve as foundational scaffolding to generate minimal models for other stimuli-responsive behaviors grounded in similar physical principles. 
Secondly, the simplicity and transparency of FHP physics make it suitable for benchmarking and hypothesis testing in atomistic and coarse-grained simulation, and developing rigorous routines for bottom-up coarse-graining.\cite{R:2021_Dhamankar_Chemically,R:2022_Jin_Bottom,R:2023_Noid_Perspective:}
Thirdly, while the present work emphasizes fundamental conceptual physics and fitting to select experimental data, it is desirable to establish a connection or mapping between the physical framework of FHP with chemically realistic systems, such as phase-separating disordered proteins and RNAs.\cite{R:2023_Rekhi_Role,R:2023_Wadsworth_RNAs,R:2024_An_Active}
Achieving quantitative consistency in some systems may require extensions to consider neglected factors, such as compressibility, which has been duly considered and effective elsewhere.}

\begin{acknowledgments}
S.D. and M.A.W. acknowledge support from the National Science Foundation under Grant No. 2237470. Simulations were performed using resources from Princeton Research Computing at Princeton University, which is a consortium led by the Princeton Institute for Computational Science and Engineering (PICSciE) and Office of Information Technology’s Research Computing.
\end{acknowledgments}

\bibliography{master}

\end{document}


\vspace*{-3cm}
    {\let\newpage\relax\maketitle}
\vspace*{-2cm}
\noindent \textbf{{\large Contents}}

\footnotesize{
I. Supplemental Text:
\begin{enumerate}[label=S\arabic*.]
    \item Derivation of the mean field free energy.
    \item Constructing another Hamiltonian from energetic and combinatorial parameters. 
    \item Representative phase behavior for distinct $p_v$ and $p_{\Omega}$.
    \item Monte Carlo simulation details.
    \item Characterization of monomer solvation environment. 
    \item Comparison of critical temperatures to single-chain transition markers.
    \item Parameters to reproduce experimental coexistence curves. 
\end{enumerate}

II. Supplemental Figures:
\setlist[enumerate,1]{leftmargin=2cm}
\begin{enumerate}[label=Fig. S\arabic*.]
    \item An extensive exploration of parameter space as $p_{\Omega}, p_v$ is varied. 
    \item Decomposition of heat capacity per its individual energetic components. 
    \item Relative distribution of contacts of a monomer with other monomers and solvent particles over temperature for the simulations from \Rone, \Rtwo, \Rthree.
    \item Examination of microscopic signatures of conformational transitions observed in different energetic regimes. 
\end{enumerate}

III. Supplemental Tables:
\begin{enumerate}[label=Table S\arabic*.]
    \item Description of the moves performed in the Monte Carlo simulation. The moves performed in the simulation perturb the location and/or the orientation of a certain set of particles on the lattice.
    \item Numerical verification ofthe enhanced sampling scheme.
    \item Comparison of the analytically computed critical temperature $T^*$ with the numerically approximated $T^{\text{conf}}$ and $T^{\text{max}}$ from simulation from \Rone, \Rtwo, and \Rthree.
    \item Parameters for FHP models shown in Fig. 4 of the main text. 
\end{enumerate}
\vspace{0.1cm}
}
\newpage 
\section{Derivation of the mean field free energy}\label{sec:SIderiv}

Consider a lattice with $N$ sites with a coordination number of $z$, populated with $m$ polymer chains of degree of polymerization $M$, and $N_s$ solvent molecules. 
All lattice sites are populated with either monomer segments or solvent molecules. 
The polymer has a volume fraction of $\varphi _p$, such that the solvent volume fraction is $\varphi _s = 1-\varphi _p$. 
Each particle is assigned an orientation $\hat{\sigma}$ that can affect energetic interactions with other particles. 

We simply consider interactions to be either \textit{aligned} or \textit{misaligned} according to the particle orientations. 
The magnitude of an \textit{aligned} interaction between particles of species $i,j$ is $\epsilon _{ij} ^{\parallel}$; the magnitude of a \textit{misaligned} interaction is $\epsilon _{ij} ^{\nparallel}$. 
Depending on the Hamiltonian of the system, a particle can engage in aligned interactions with only a subset of its $z$ neighbors, while the neighboring particles outside of this set can only engage in misaligned interactions. 
We define the fraction of neighbors that can engage in aligned interactions as  $p_v$. 
A Hamiltonian that allows for aligned interactions with all $z$ neighbors has $p_v=1$, while a Hamiltonian that allows for no aligned interactions has $p_v =0$. 

 Suppose that particles adopt discrete orientations from the set $S$, and the set has $z$ elements. 
 The probability that two neighboring particles will be aligned based on uniform random sampling is 
\begin{align}
    p _{\Omega} = \frac{\mathlarger{\sum _{\Hat{\sigma} _i \in S}} \, \mathlarger{\sum_{\Hat{\sigma} _j \in S}} \delta _{\Hat{\sigma} _i\Hat{\sigma} _j}}{\mathlarger{\sum _{\Hat{\sigma} _i \in S }} \, \mathlarger{\sum _{\Hat{\sigma} _j \in S}} 1}= \frac{n^{\parallel}}{z^2},
\end{align}
where $n^{\parallel}$ is result of the summations in the numerator.
In this fashion, $p_\Omega $ would also be the probability of an aligned interaction if $\epsilon _{ij} ^{\parallel}$ = $\epsilon _{ij} ^{\nparallel}$.
Under Boltzmann statistics at inverse temperature $(\beta = 1/k_{\text{B}}T)$, the expected interaction energy between two particles that can exhibit an aligned interaction is then 
\begin{align}
    \left\langle \epsilon _{ij}\right\rangle  &= \mathlarger{\sum _{\gamma \in \{\parallel, \nparallel \}}} \epsilon _{ij} ^{\gamma} p_{\Omega} ^{\gamma} \exp \left(-\beta \epsilon _{ij} ^{\gamma} \right) \\
    &= \frac{p_{\Omega} \exp \left(-\beta \epsilon _{ij} ^{\parallel} \right)\epsilon _{ij} ^{\parallel} + (1-p_{\Omega}) \exp \left(-\beta \epsilon _{ij} ^{\nparallel} \right)\epsilon _{ij} ^{\nparallel} }{p_{\Omega} \exp \left(-\beta \epsilon _{ij} ^{\parallel} \right)+(1-p_{\Omega}) \exp \left(-\beta \epsilon _{ij} ^{\nparallel}\right)},
\end{align}
such that the mean interaction between two particles is
\begin{align}
    w_{ij} &= p_v\left\langle \epsilon _{ij}\right\rangle  + (1-p_v) \epsilon _{ij} ^{\nparallel} \\
    &= 
    \epsilon _{ij} ^{\nparallel} + p_v \biggl(\mathlarger{\sum _{\gamma \in \{\parallel, \nparallel \}}} \epsilon _{ij} ^{\gamma} p_{\Omega} ^{\gamma} \exp (-\beta \epsilon _{ij} ^{\gamma}) - \epsilon _{ij} ^{\nparallel} \biggr) 
\end{align}
This ultimately gives the energy of mixing as 
\begin{align}
    \Delta U _{\text{mix}} &= k_{\text{B}}TN_1 \varphi _2 \chi _{\text{FHP}} (T) \\ 
    \chi _{\text{FHP}} (T) &= \frac{(z-2)}{k_{\text{B}}T} \left( w_{\text{ms}} - \frac{1}{2} \left( w_{\text{mm}} + w_{\text{ss}} \right) \right) 
\end{align}
where `m' and `s' stands for monomer species and solvent species, respectively. 

To recast $\chi _{\text{FHP}} (T)$ as a perturbation from a standard Flory-Huggins $\chi$ parameter generated by purely misaligned $(\nparallel)$ interactions,
we divide the numerator and denominator of $\langle \epsilon _{ij} \rangle$ by $p_{\Omega}\exp \left(-\beta \epsilon _{ij} ^{\nparallel}\right)$ and use $\epsilon _{ij} ^{\parallel} \leftarrow \Delta _{ij} + \epsilon _{ij} ^{\nparallel}$:
\begin{align}
    \chi _{\text{FHP}} (T) &= \frac{(z-2)}{k_{\text{B}}T} \left( w_{\text{ms}} - \frac{1}{2}\left( w_{\text{mm}} + w_{\text{ss}}\right)\right) \nonumber \\ 
    &= \frac{(z-2)}{k_{\text{B}}T}  \left( p_v \left(  \frac{\exp\left( -\beta \Delta _{\text{ms}} \right) \Delta _{\text{ms}} }{ \exp\left( -\beta \Delta _{\text{ms}} \right) + r } - \frac{1}{2} \mathlarger{\sum _{i\in \{\text{m,s} \}}}  \frac{\exp\left( -\beta \Delta _{ii} \right) \Delta _{ii} }{ \exp\left( -\beta \Delta _{ii} \right) + r } \right) + \left( \epsilon _{\text{ms}} ^{\nparallel} - \frac{1}{2} \left( \epsilon _{\text{mm}} ^{\nparallel} + \epsilon _{\text{ss}} ^{\nparallel}\right) \right) \right) \nonumber \\
    &= \frac{(z-2)}{k_{\text{B}}T} \left( \epsilon _{\text{ms}} ^{\nparallel} - \frac{1}{2} \left( \epsilon _{\text{mm}} ^{\nparallel} + \epsilon _{\text{ss}} ^{\nparallel}\right) + p_v \left( \frac{\Delta _{\text{ms}} }{ 1 + r\exp(\beta \Delta _{\text{ms}}) }  - \frac{1}{2} \mathlarger{\sum _{i\in \{\text{m,s} \}}} \left( \frac{\Delta _{ii} }{ 1 + r\exp(\beta \Delta _{ii}) } \right) \right) \right) \nonumber \\
    &= \frac{(z-2)}{k_{\text{B}}T} \left( \epsilon _{\text{ms}} ^{\nparallel} - \frac{1}{2} \left( \epsilon _{\text{mm}} ^{\nparallel} + \epsilon _{\text{ss}} ^{\nparallel}\right)\right) + \frac{(z-2)}{k_{\text{B}}T}\left( p_v \left( \Tilde{\Delta} _{\text{ms}}(T) - \frac{1}{2}\left(\Tilde{\Delta} _{\text{mm}}(T) + \Tilde{\Delta} _{\text{ss}}(T) \right) \right) \right) \nonumber \\
    &= \chi _{\text{FH}}(T) + \Tilde{\chi} (T) \label{eq:chi_FHP}
\end{align}
where we define
\begin{equation}
    \Tilde{\Delta} _{ij}(T) = \frac{\Delta _{ij}}{1+r\exp \left(\Delta _{ij} /k_{\text{B}}T\right)}
\end{equation}
for notational convenience.

Mixing does not affect the accessible orientations of any particle species, such that the 
the entropy of mixing follows that of Flory-Huggins theory: 
\begin{align}
    \Delta S _ {\text{mix}} = -k_{\text{B}} (N_1 \ln \phi _1 + N_2 \ln \phi _2)
\end{align}

Therefore, the free energy of mixing per particle is given by
\begin{align}
    \Delta \overline{F} _{\text{mix}} &= \Delta \overline{U}_{\text{mix}} - T\Delta \overline{S}_{\text{mix}} \nonumber \\
    \implies \beta \Delta \overline{F} _{\text{mix}} &= \frac{\varphi}{M} \ln \varphi + (1-\varphi)\ln (1-\varphi) + \chi _{\text{FHP}}(T)\varphi(1-\varphi) 
\end{align}

\section{Constructing a Hamiltonian from mean-field analysis}
A key feature of this framework is the direct translation of the parameters provided in the framework to a system of particles on a lattice than interact with a tractable Hamiltonian. This can then be simulated to investigate the physics at a more foundational, microscopic level. 

After obtained mean-field parameters, such as those acquired through parameter fitting to experimental data, there exists a principled approach to derive a Hamiltonian for subsequent simulations.

Example 1. If the mean-field analysis is satisfied if $p_v = 1$, then this implies that the particles in the system can engage in aligned interactions with all of their neighbors. Furthermore, if the analysis also supports $p_{\Omega} = 0.25$, then for two particles that can engage in aligned interactions, only $25\%$ of all possible orientational states lead to an aligned interaction. Therefore, one has to construct a Hamiltonian where the particle of interest interacts with all of its neighbors, and can be aligned with all of its neighbors with a probability of $0.25$ for every neighbor. 
One possible formulation of such a Hamiltonian is
\begin{equation}
\Lambda_{\text{corr}}(i,j)  = \Theta (\hat{\sigma} _i \cdot \hat{\sigma} _j - \delta), 
\end{equation}
where $\delta = \cos 1$ ($\cos 1$ was obtained using a simple numerical testing procedure), and $\Theta$ is the Heaviside theta function. Given these parameters are the possible 26 orientations on a cubic lattice, all the physical conditions elucidated by the mean-field analysis have been met (see Fig. 1(b)). This style of interactions is referred to as a correlation network. 

Example 2. If the mean-field analysis is satisfied if $p_v = 3/8$, then this implies that the particles in the system can engage in aligned interactions with only $25\%$ of their neighbors. Furthermore, if the analysis supports $p_{\Omega} = 0.35$, then for two particles that can engage in aligned interactions, only $35\%$ of all possible orientational states lead to an aligned interaction. Therefore, one has to construct a Hamiltonian where the particle of interest interacts with only $3/8^{\text{th}}$ of its neighbors and can only be aligned with those neighbors with a probability of $0.25$. 
One possible for formulation of such a Hamiltonian is
\begin{equation}\label{eq:LambdaLock}
    \Lambda_{\text{lock}} (i,j) = \Theta ( \delta - \left[ \arccos \left( \hat{r}_{ji} \cdot \hat{\sigma} _i \right) + \arccos \left( \hat{r}_{ij} \cdot \hat{\sigma} _j \right) \right] ) 
\end{equation}
where $\hat{r} _{ij} = \frac{\vec{r} _i - \vec{r} _j}{|\vec{r} _i - \vec{r} _j|}$, $\delta = \pi/2$ ($\pi/2$ was obtained by simple numerical testing), and $\Theta$ is the Heaviside theta function. Given these parameters are the possible 26 orientations on a cubic lattice, all the physical conditions elucidated by the mean-field analysis have been met (see Fig. 1(c)). This style of interactions is referred to as a selective locking.

\subsection{Determination of the critical point and $\chi^*$}\label{sec:SIcrit}
Whether a given polymer solution can be in a homogeneous or phase-separated state at a given temperature is determined by the value of \chieff relative to $\chi ^*$-- the value of $\chi$ at the critical point. 
If $\chieff < \chi ^*$, the solution must be homogeneous at all compositions; if $\chieff > \chi ^*$, the solution will phase-separate.

The critical point of the polymer is given by
\begin{align}
    \left. \frac{\partial ^2 \beta \Delta \overline{F}_{\text{mix}}}{\partial \phi ^2} \right|_{T} &= \frac{1}{1-\varphi} + \frac{1}{M\varphi} -2 \chi = 0 \\ 
    \left. \frac{\partial ^3 \beta \Delta \overline{F}_{\text{mix}}}{\partial \phi ^3} \right|_{T} &= \frac{1}{(1-\varphi)^2} - \frac{1}{M\varphi ^2} = 0 .
\end{align}
Solving the above equations then yields
\begin{align}
    \varphi ^* &= \frac{-1\pm \sqrt{M}}{-1+M} \\
    \chi ^* &= \frac{\left(\sqrt{M} \pm 1\right)^2}{2M} \label{eqn:critical-chi}
\end{align}
For $M=32$, the physically meaningful solutions are $\varphi ^* \approx 0.150$ and $\chi ^* \approx 0.692$. 

\section{Representative phase behavior for distinct $p_v$ and $p_{\Omega}$}\label{sec:SIexplore}
\begin{figure}[H]
\centering
\includegraphics[width=0.85\textwidth,keepaspectratio]{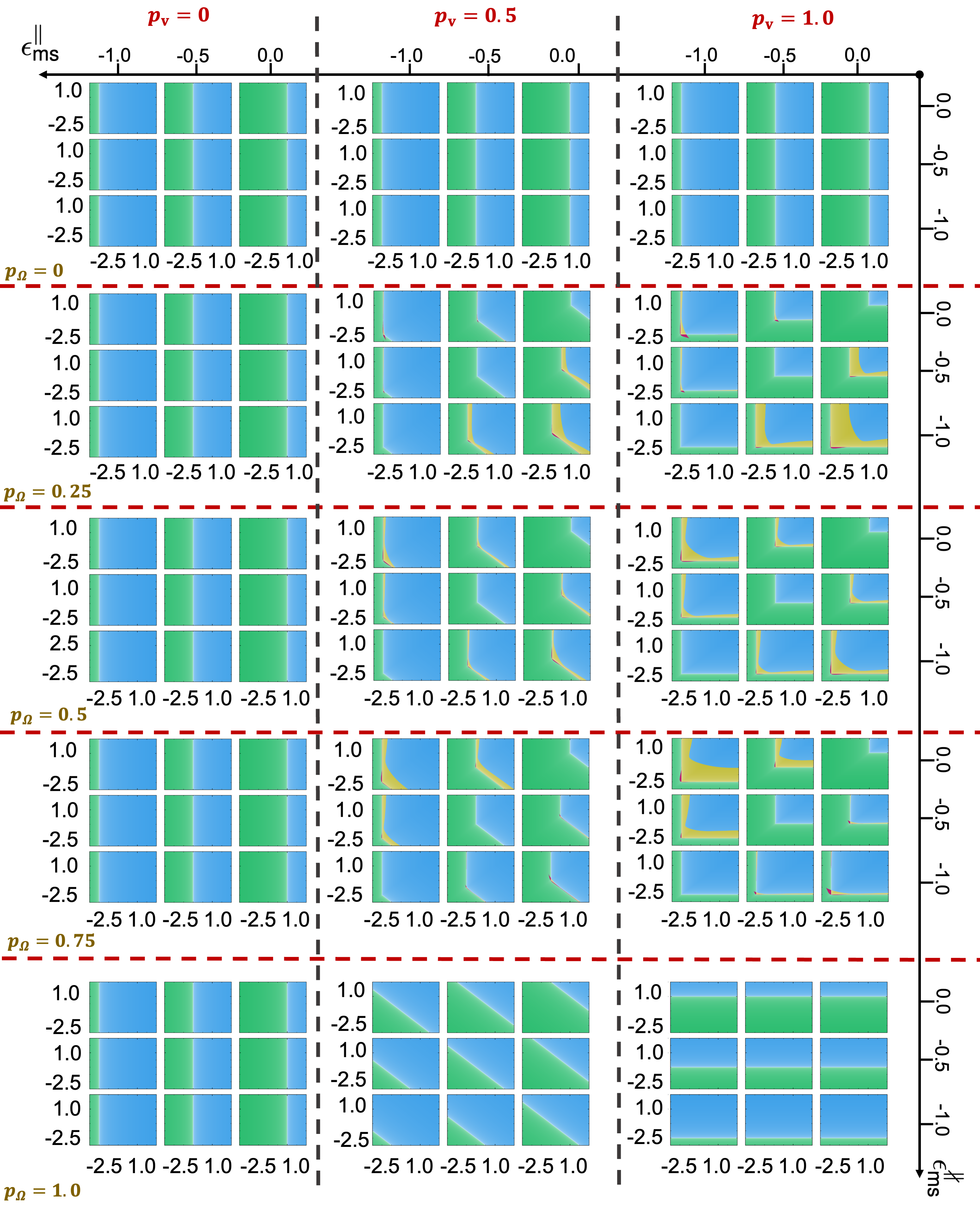}
\caption{An exploration of parameter space as $p_{\Omega}, p_v$ is varied. For each tuple of $(p_{\Omega}, p_v)$, $\Emsa, \Emsn$ are either -1, -0.5, or 0, while $\Emma, \Emmn$ varies continuously over $[-2.5, 1]$.}\label{SI_Fig1:Param_exploration}
\end{figure}

To more broadly characterize phase behavior as a function of the FHP parameter-space, we also explore implications of $p_v$ and $p_{\Omega}$. 
Specifically, we examine systems for which $p_v$ is 0.0, 0.5, or 1.0 and $p_{\Omega}$ is 0.0, 0.25, 0.5, 0.75, or  1.0. 
The results are shown in Fig. {\ref{SI_Fig1:Param_exploration}}

When interactions are purely isotropic ($p_v=0, p_{\Omega}=0 \text{ or }1$), there are no regions of $\mathcal{R}_2$ or $\mathcal{R}_3'$.
This can be anticipated by introducing these limits to Eq. \eqref{eq:chi_FHP}, we observe the following limiting behaviors
\begin{align}
    \lim_{T\to 0^+,p_v \to 0} \chieff = \frac{z-2}{k_{\text{B}}T}\left(\Emsn - \frac{1}{2}\left(\Emmn + \Essn \right) \right)
\end{align}
and 
\begin{align}
    \lim_{T\to 0^+,p_{\Omega} \to 0} \chieff = \frac{z-2}{k_{\text{B}}T}\left(\Emsn - \frac{1}{2}\left(\Emmn + \Essn \right) \right)
\end{align}
and 
\begin{align}
    \lim_{T\to 0^+,p_{\Omega} \to 1} \chieff = \frac{z-2}{k_{\text{B}}T}\left(p_v \left(\Emsa - \frac{1}{2}\left( \Emma + \Essa \right) \right) + (1-p_v)(\Emsn - \frac{1}{2}\left( \Emmn + \Essn \right) \right) 
\end{align}

Therefore, when $p_{\Omega}$ or $p_v$ is zero, the boundary between {\Rzero} and {\Rone} is vertical. 
As $p_v$ is increased with $p_{\Omega}$ fixed at $1$, the boundary starts to rotate counterclockwise, culminating in a horizontal configuration as $p_v \to 1$. 
As $p_v$ is increased from $0$ to $1$, when $p_{\Omega}$ does not tend to $0$ or $1$, the boundary starts to take more of an $L$-shape. 
For $p_v=1$, the boundary parallel to the $\Emmn$ axis has a negative slope for $0<p_v<1$, while the boundary parallel to the $\Emma$ axis remains vertical for all $p_v$. 

\section{Monte Carlo simulation details}\label{sec:SImontecarlo}

All simulations feature a $34\times 34\times 34$ simple cubic lattice with periodic boundaries; lattice sites are occupied by solvent particles and a single polymer chain with $M=32$.
We use $z=26$ (accounting for interactions between nearest, next-nearest, and next-next-nearest neighboring sites) and restrict particle orientations to direct towards the 26 neighboring sites (Fig. 1a of main text). 
MC moves for solvent particles include solvent-orientation exchanges and group-level orientation perturbations.
MC moves for the polymer include conventional end-rotation, reptation, orientation perturbations, and chain regrowth with orientation perturbations. The complete set of moves utilized in our simulations are shown in Table {\ref{tab_SI:mcmoves}}; moves are selected with equal probability.  
To enhance sampling efficiency, chain-regrowth moves are implemented using a Rosenbluth sampling scheme.
All simulations consist of $10^8$ MC moves with the latter half used for analysis; averages are obtained from 30 independent trajectories with all errors representing the standard error of the mean.

\begin{table}[H]
    \centering
        \caption{Description of the moves performed in the Monte Carlo simulation. The moves performed in the simulation perturb the location and/or the orientation of a certain set of particles on the lattice.  }
    \resizebox{\textwidth}{!}{
    \normalsize
    \begin{tabular}{|c|c|c|}
         \hline
         & Move type & Move description  \\
         \hline
         \hline
         1 & \makecell[l]{End rotation} & \makecell[l]{The monomer segment at the ends of the polymer chain, $m_1$ and $m_M$, are uniformly \\ chosen and swapped with a neighboring solvent particle of $m_2$ or $m_{M-1}$, respectively.} \\
         \hline
         2 & \makecell[l]{Reptation\\`slithering snake'}  & \makecell[l]{Uniformly picking an end of the polymer ($m_1$ or $m_M$), the polymer is perturbed/pushed in a \\ particular direction.  This move is accepted only if the polymer is `chasing itself' \\($m_1$ moving to the location of $m_M$, or vice versa), or if the polymer is displacing a solvent particle.} \\
         \hline
         3 & \makecell[l]{Chain regrowth \\ with orientation \\ perturbation} & \makecell[l]{A monomer index $i$ is chosen uniformly from index $2$ to $M-1$. \\If $|M-i| < |i-1|$, then the monomers $m_{i+1}, m_{i+2},\ldots,m_M$ have their locations \\and orientation perturbed; \\if $|M-i| > |i-1|$, then the monomers $m_1, m_2, \ldots, m_{i-1}$ i.e. \\ have their locations and orientation perturbed; \\if $|M-i| = |i-1|$, one of the aforementioned sets \\$\left(\{ m_1, m_2,\ldots, m_{i-1}\}, \{m_{i+1}, m_{i+2}, \ldots, m_M\} \right)$  of monomers is chosen uniformly. } \\
         \hline
         4 & \makecell[l]{Chain regrowth \\ without orien-\\tation perturbation} & \makecell[l]{A monomer index $i$ is chosen uniformly from index $2$ to $M-1$. \\If $|M-i| < |i-1|$, then the monomers $m_{i+1}, m_{i+2},\ldots,m_M$ have their locations perturbed;\\ if $|M-i| > |i-1|$, then the monomers $m_1, m_2, \ldots, m_{i-1}$ have their locations perturbed;\\ if $|M-i| = |i-1|$, one of the aforementioned sets \\$\left(\{ m_1, m_2,\ldots, m_{i-1}\}, \{m_{i+1}, m_{i+2}, \ldots, m_M\} \right)$ of monomers is chosen uniformly. } \\
         \hline
         5 & \makecell[l]{Polymer orientation\\ perturbation} & \makecell[l]{The orientations of a random set of monomers out of $M$ is perturbed.} \\
         \hline 
         6 & \makecell[l]{Solvent orientation\\ perturbation} & \makecell[l]{Out of a total of $(L/2)^3$ solvent particles in the simulation cell, $n$ random solvent\\ particles are selected uniformly at random and their orientations are perturbed.}\\
         \hline
         7 & \makecell[l]{Solvent exchange} & \makecell[l]{Two solvent particles are randomly chosen and their locations are swapped.}
         \\
         \hline
         8 & \makecell[l]{Solvent exchange\\with orientation\\ perturbation} & \makecell[l]{Two solvent particles are randomly chosen and their locations are swapped\\ and orientations are randomly perturbed.} \\ 
         \hline 
         9 & \makecell[l]{Solvation shell \\ perturbation} & \makecell[l]{Solvent particles that are in contact with the polymer have their orientation randomly perturbed.} \\
         \hline 
         10 & \makecell[l]{Solvation shell\\ exchange} & \makecell[l]{A random solvent particle from the solvation shell and a random particle\\ in the simulation cell is chosen and their locations are swapped. } \\ 
         \hline 
         11 & \makecell[l]{Solvation shell\\ exchange with \\orientation perturbation} & \makecell[l]{A random solvent particle from the solvation shell and a random particle in the simulation cell\\ is chosen and their locations are swapped and their orientations are perturbed. } \\
        \hline 
    \end{tabular}}
    \label{tab_SI:mcmoves}
\end{table}

\subsection{Chain regrowth with orientation perturbation}
In addition to conventional Monte Carlo (MC) moves, our simulations also employ chain-regrowth moves that simultaneously perturb particle orientations  and alter the polymer configuration. 
The regrowth process begins by selecting a monomer $m_i$ at index $i$, which serves as the hinge point.
Then, a new state of the chain is generated by perturbing the locations and orientations of monomers $m_{i+1}$ through $m_M$ step-by-step. During this process, bond constraints are relaxed, and the energy is calculated solely based on non-bonded interactions, but the polymer is restored to a well-connected state by the final step to preserve detailed balance and connectivity constraints.

To derive the appropriate acceptance criterion when subjected to Rosenbluth sampling, we write the energy of the system as $ U ( r^{M}_{(k)}, r_{k}, \sigma ^M _{(k)}, \sigma _{k} ) $, where
\begin{enumerate}
    \item $r_{k}$, $\sigma_{k}$ represent the location and orientation of monomer $m_k$, respectively.
    \item $r^M_{(k)}$ represents the positions of all $M$ monomers, excluding $m_{k}$.
    \item $\sigma^M_{(k)}$ represents the orientations of all $M$ monomers, excluding $m_{k}$.
\end{enumerate}
The proposed final state of the system is indicated by $n$ (new), and the initial state with $o$ (old).

We first consider a set of $k<z$ trial locations of the monomer, $\{\ell\}_{i+1}$, that are the nearest-neighbors of $m_i$. If $m_{i+1}$ swaps positions with the nearest-neighbors of particle $m_i$, the swap is only permitted if the neighboring particles are solvent particles or occupied by $m_j$ such that $j>i+1$. If the location is occupied by a $m_j$ such that $j<i+1$, its energy will be $U( r^{M}_{(i+1)}, r_{i+1}, \sigma ^M _{(i+1)}, \sigma _{i+1}) = \infty$. 

In the following, we denote local configurations and orientations associated with the present state with a double prime ($''$) and a proposed state with a single prime ($'$). To calculate the forward rate, we first calculate the Boltzmann sum, $w_1(i+1)$, of these trial locations:
\begin{align}
    w_1(i+1) = \sum _{ \ell \in \{\ell \}_{i+1} } \exp\left(-\beta U\left(r^M _{(i+1)},\ell,\sigma ^M_{(i+1)}, \sigma _{i+1}\right) \right) 
\end{align}
Of the $k$ trial locations, we select one configuration $\ell'$ with probability 
\begin{align}
    p_1(i+1) = \frac{\exp ( -\beta U ( r^M _{(i+1)},\ell ',\sigma ^M_{(i+1)}, \sigma _{i+1} ) )}{w _i \left(i+1 \right)}.
\end{align}
Following this selection, we then sample orientations of $m_{i+1}$ from the set of possible of orientations $\{ o \} _{i+1}$. To calculate the forward rate for perturbing orientation of $m_{i+1}$ from $q''$ to $q'$, we  calculate a Boltzmann sum $w _2(i+1)$ of these trial orientations 
\begin{align}
    w_2(i+1) = \sum _{q \in \{ o\} _{i+1} } \exp \left(-\beta U \left( r^M _{(i+1)}, \ell ', \sigma ^M_{(i+1)}, q \right) \right) 
\end{align}
and select one orientation $q'$ with probability 
\begin{align}
    p_2 (i+1) = \frac{\exp (-\beta U ( r^M _{(i+1)}, \ell ', \sigma ^M_{(i+1)}, q' ) ) }{w_2(i+1)}.
\end{align}
The probability of selecting both $\ell '$ with spin $q'$ for monomer $m_{i+1}$ is then
\begin{align}
    W^{\text{forw}}(i+1) = p _1(i+1)p _2(i+1) 
\end{align}
This procedure is repeated for $m_{i+2}, m_{i+3}, ..., m_{N}$ to regrow the polymer chain to a new configuration $n$. 
The complete regrowth to state $n$ follows 
\begin{align}
    W ^{\text{forw}}(o\rightarrow n) = \prod _{j=i+1} ^N W^ {\text{forw}} (j)
\end{align}

The backward rate $W ^{\text{back}}(n\rightarrow o)$ can be computed similarly by gradually perturbing state $n$ into state $o$, starting from $m_{i+1}$ and resampling its locations and spins from $\ell '$ to $\ell ''$ and $q'$ to $q''$. In the reverse move, the total Boltzmann sum $w _i'(i+1)$ over trial locations is  
\begin{align}
    w_1 '(i+1) = \sum _{ \ell \in \{\ell \}_{i+1} } \exp ( -\beta U ( r^M _{(i+1)},\ell,\sigma ^M_{(i+1)}, \sigma _{i+1} ) )
\end{align}
The probability of choosing $\ell ''$ is 
\begin{align}
    p_1 '(i+1) = \frac{\exp ( -\beta U ( r^M _{(i+1)},\ell '',\sigma ^M_{(i+1)}, \sigma _{i+1} ) )}{w _1'\left(i+1 \right)}
\end{align}
We again sample the orientations of $m_{i+1}$($\sigma _{i+1}$) from the set of possible of orientations $\{ o \} _{i+1}$ and find the probability of reverting the spin back to $q''$. The Boltzmann sum here is 
\begin{align}
    w _2' (i+1) = \sum _{q \in \{ o\} _{i+1} } \exp (-\beta U ( r^M _{(i+1)}, \ell '', \sigma ^M_{(i+1)}, q ) ) 
\end{align}
and the orientation $q''$ is selected with probability 
\begin{align}
    p _2 '(i+1) = \frac{\exp \left(-\beta U \left( r^M _{(i+1)}, \ell '', \sigma ^M_{(i+1)}, q '' \right) \right) }{w _2 '(i+1)}
\end{align}

Therefore, for monomer $m_{i+1}$ we have 
\begin{align}
    W^{\text{back}}(i+1) = p _1 '(i+1)p _2 '(i+1), 
\end{align}
the total weight of the backward move is
\begin{align}
    W ^{\text{back}}(n\rightarrow o) = \prod _{j=i+1} ^N W^ {\text{back}} (j) .
\end{align}

This results in an acceptance criterion for the chain regrowth that follows the traditional form but with modified internal weights: 
\begin{align}
    \text{acc} (o\rightarrow n) = \min \left(1, \exp(-\beta (U(n) - U(o))) \cdot \frac{W^{\text{back}}(n\rightarrow o)}{W^{\text{forw}}(o \rightarrow n)} \right)
\end{align}
where $U(n)$ is the energy of the new state and $U(o)$ is the energy of the old state. 

We note that the above procedure can be reconfigured to incorporate both location and orientation perturbations in trial configurations, rather than making orientation-sampling conditional on the location-sampling. This may ultimately be more efficient but was not required for the present study. 

\subsection{Numerical verification of detailed balance}

To verify the Rosenbluth sampling scheme, we compared empirically obtained transition rates using our sampling scheme with the corresponding state probabilities determined by the Boltzmann distribution.
In particular, expectations from Boltzmann statistics are given by
\begin{align}
    R_{\text{true}} = \frac{\rho(n)}{\rho(o)} = \frac{\exp (-\beta U(n))}{\exp (-\beta U(o))} = \exp (-\beta(U(n)-U(o)))
\end{align}
This is compared with an empirical rate based on evaluating $(N(o\rightarrow n)$, the number of times the system accepts moves to reach state $n$ from state $o$, and $(N(n\rightarrow o))$, the number of times the system accepts a move to reach state $o$ from state $n$. Then, transition rate of going from state $o$ to state $n$ is given by
\begin{align}
    R_{\text{meas}} = \frac{N(o\rightarrow n)}{N(n\rightarrow o)}
\end{align}
Table {\ref{tab:my_label}} summarizes the results. In general, the Rosenbluth sampling scheme produces the correct population distribution within statistical error as intended. Larger errors and uncertainties are incurred when there are more extreme energy differences in Boltzmann weights.

\begin{table}[H]
    \caption{Numerical tests of the Rosenbluth sampling scheme. Two tests were performed: one in which the new state $(n)$ involves a change in location and spin of just one monomer unit (first nine entries) and one in which the new state $(n)$ involves a change in location and spin of two monomers (remaining entries). Tests were performed at three (reduced) temperatures $(T=1,2,3)$. For statistics,  three independent trials each with $10^7$ moves were run at each temperature.  The percent error of the measured transition $R_{\text{meas}}$ rate $\left(\left(R_{\text{meas}}-R_{\text{truth}}\right)/R_{\text{truth}}\cdot100 \right)$ from the analytically determined transition rate $R_{\text{truth}}$ is reported along with the standard error.}
    \label{tab:my_label}
    \centering
    \resizebox{0.8\textwidth}{!}{
    \begin{tabular}{|c|c|c|c|c|c|}
         \hline
         $T$ & Number of moves & Test locations $(k)$ & $R_{\text{truth}}$& $R_{\text{meas}}$ & \% error \\
         \hline
         \multicolumn{6}{|c|}{Tests where transition involved a single monomer changing location and spin.} \\
         \hline
         1&  $10^7$&  5&  ~7.3891 &  ~7.395  $\pm$ 0.004 &   0.08 \\
         2&  $10^7$&  5&  ~2.7183 &  ~2.722  $\pm$ 0.002 &   0.13  \\
         3&  $10^7$&  5&  ~1.9477 &  ~1.950  $\pm$ 0.002 &  -0.10\\
         \hline
         1&  $10^7$&  5&  ~7.3891&  ~7.41 $\pm$ 0.03 &   0.35   \\
         2&  $10^7$&  5&  ~2.7183&  ~2.71 $\pm$ 0.01 &  -0.26   \\
         3&  $10^7$&  5&  ~1.9477&  ~1.940 $\pm$ 0.007 &  -0.41  \\
         \hline 
         \multicolumn{6}{|c|}{Tests where transition involved two monomers changing location and spin.} \\
         \hline
         1& $10^7$& 5& 54.5982 & 53.1 $\pm$ 0.7 & -2.67  \\
         2& $10^7$& 5& ~7.3891  & ~7.34  $\pm$ 0.04 & -0.70  \\
         3& $10^7$& 5& ~3.7936  & ~3.79  $\pm$ 0.08 & -0.08  \\ 
         \hline
         1& $10^7$& 5 & 54.5982 & 56 $\pm$ 1 & 2.4  \\
         2& $10^7$& 5 & ~7.3891  & ~7.40  $\pm$ 0.02 & 0.09 \\
         3& $10^7$& 5 & ~3.7936  & ~3.77   $\pm$ 0.02 & -0.74 \\          
         \hline 
    \end{tabular}}
\end{table}
 
\section{Characterization of monomer solvation environment}\label{sec:SIcompss}
\begin{figure}[H]
    \centering    
    \includegraphics[width=\textwidth,keepaspectratio]{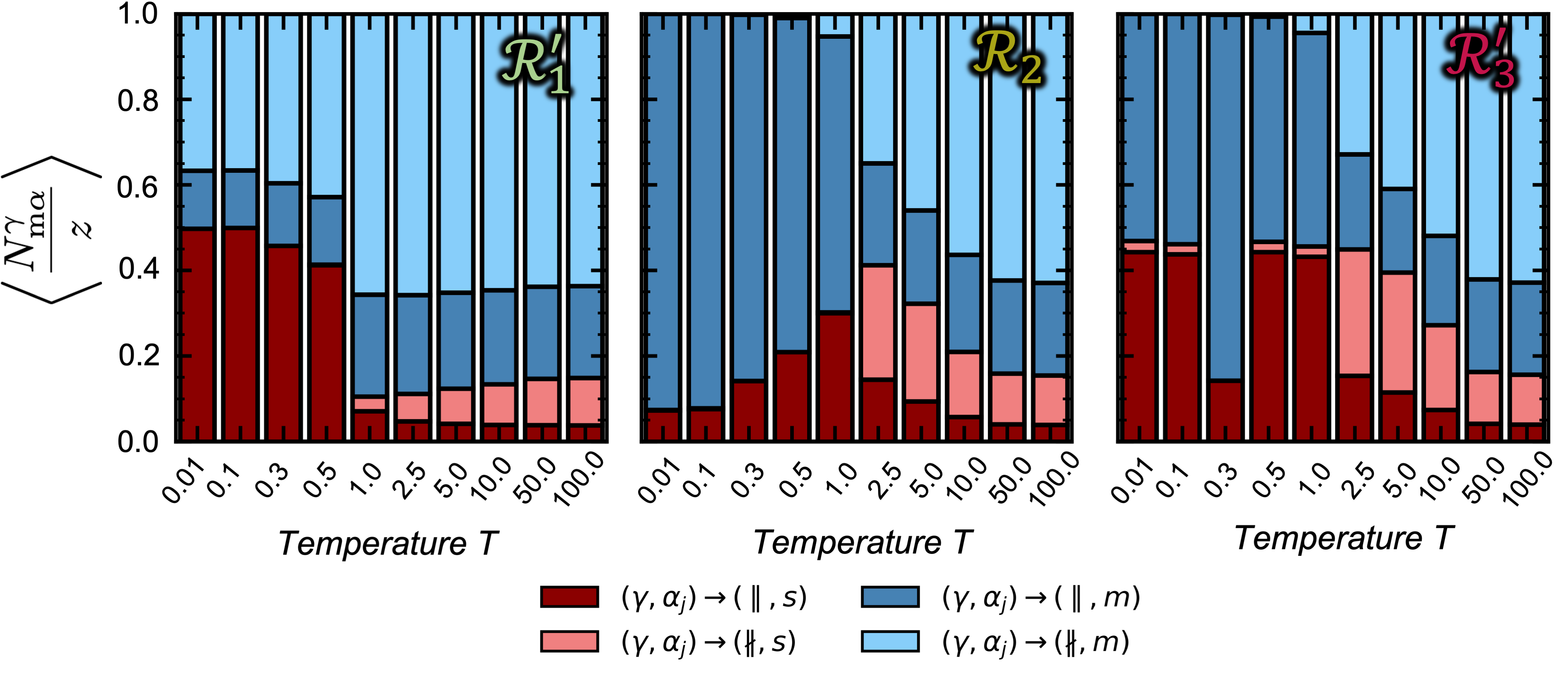}
    \caption{Relative distribution of interactions with monomers as a function of temperature. From left-to-right, data are shown for simulations using parameters linked to  \Rone, \Rtwo, and {\Rthree}. $\langle N^{\gamma} _{\text{m}\alpha}/z\rangle$ measures the fraction of interactions of style  $\gamma \in \{\parallel,\nparallel \}$ a particular monomer is engaging in with species $\alpha \in \{ \text{m, s}\}$. }\label{SI_Fig4:ContactComposition}
\end{figure}

To better understand the microscopic driving forces underlying the observed coil-globule and globule-coil transtions, we characterized the solvation environment surrounding monomer particles as a function of temperature. 
Fig. {\ref{SI_Fig4:ContactComposition}} compares the average interactions for monomer particles in parametric regimes linked to {\Rone}, {\Rtwo}, and {\Rthree} phase behavior.  We generally observe that there is growth in the number of misaligned interactions as a function of temperature preceding a major conformational transition. 
In the case of simulations from {\Rone}, the uptick in misaligned monomer interactions is gradual while many solvent interactions are retained before a major globular collapse. For simulations from {\Rtwo}, the chain first exchanges aligned monomer interactions for aligned solvent interactions before misalignment coincides with expansion to a coil.
Somewhat similar behavior is observed in {\Rthree} for the third transition. 
The brief transition to a globular state in {\Rthree} intriguingly coincides with growth in aligned monomer interactions that subsequently dissipate. 

The changes in the solvation shell of the polymer are accompanied with fluctuations in energy. We also identify key molecular-level signatures associated with the observed conformational transitions. We specifically monitor the heat capacity $\overline{C}_V$, energy fluctuations $\sigma^2_E$, and  relative proportion of misaligned to aligned interactions $N^\nparallel / N^\parallel$.
For systems from {\Rone}, the single coil-globule transition is marked by maxima in  $\overline{C}_V$ and $\sigma^2_E$ that coincide with increasing the proportion of misaligned interactions over narrow temperature range wherein $\tilde{\nu}\sim 0.5 $ (white region).
Similar observations hold for the low-$T$ transitions from a coil to a globule in {\Rtwo} and {\Rthree}, for which at least some enhancement in misaligned interactions accompanies peaks in $\overline{C}_V$. 
In addition, {\Rtwo} and {\Rthree} display a second higher-$T$ transition from globular to excluded-volume statistics that is only captured by a maximum in $\sigma^2_E$ as $N^\nparallel / N^\parallel$ plateaus towards a limiting high-$T$ value of $r = (1-p_\Omega)/p_\Omega$.
With the current data, however, we do not find perfect alignment with critical temperatures and these markers for single-chain transitions (Supplemental Materials, Section {\ref{SI_sec:markers}}).
Nevertheless, sharp peaks in $\overline{C}_V$ and/or $\sigma^2_E$ provide reasonable markers for  single-chain conformational transitions, which are ultimately driven by compositional and orientational changes in the polymer's local environment.
\begin{figure}[H]
    \centering    
    \includegraphics[width=0.7\textwidth,keepaspectratio]{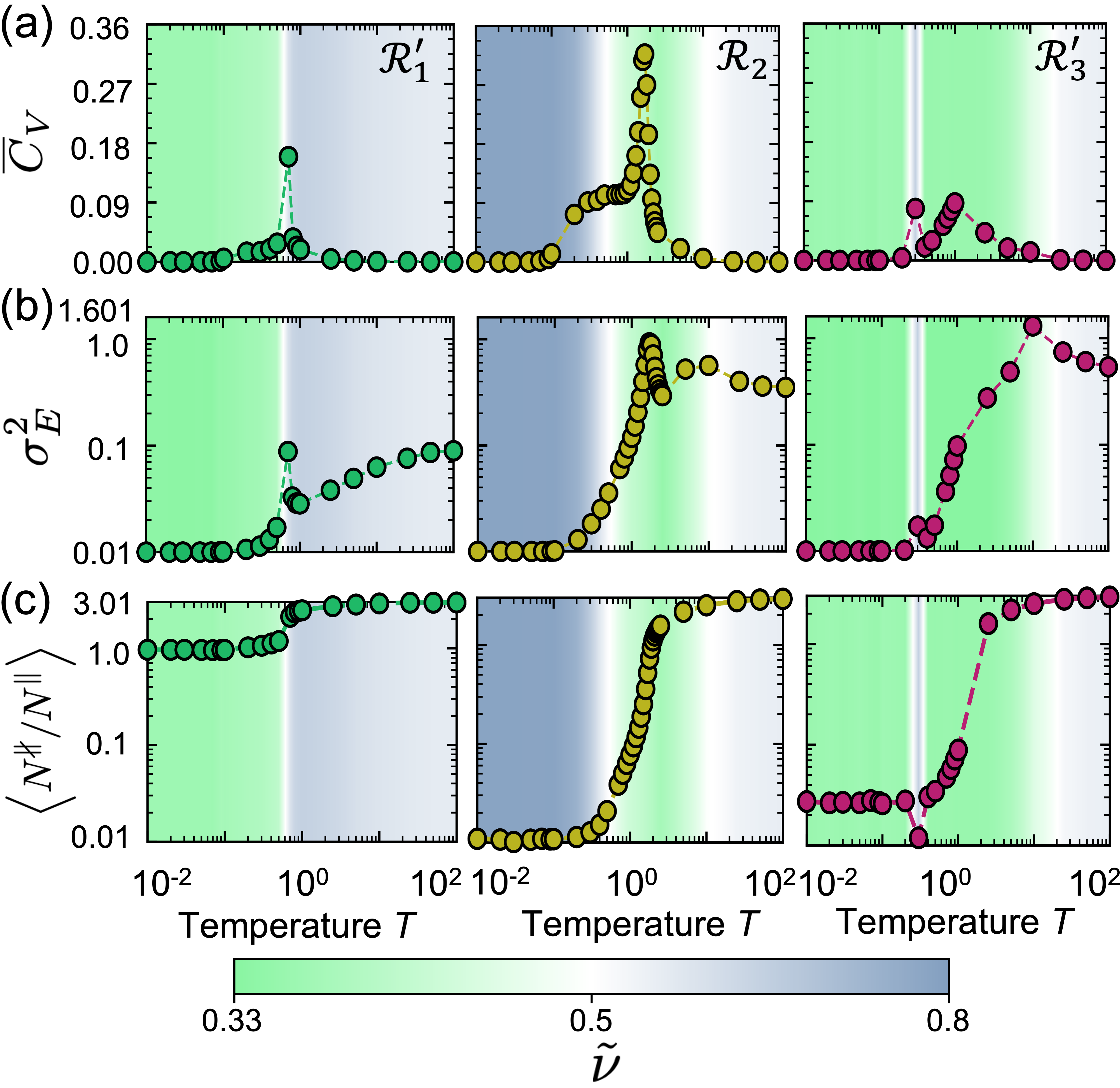}
    \caption{Examination of microscopic signatures of conformational transitions observed in different energetic regimes. Quantities include (a) heat capacity $\overline{C}_V$, (b) fluctuations in total energy $\sigma_E^2$, and (c) the average proportion of misaligned to aligned interactions $N^{\nparallel}/N^{\parallel}$. Plots in panels (b) and (c) have been shifted by $0.01$ to highlight the variation in the quantity as the conformation of the polymer changes. Each panel displays data obtained from Monte Carlo simulation of single polymer chains using parameters sets associated with {\Rone}, {\Rtwo}, and {\Rthree}, arranged left-to-right.
Results are reported for the same simulations in Fig. 4, such that the background color reflects the value $\tilde{\nu}$ at the given temperature based on interpolation. }\label{SIFig:Fluctuations}
\end{figure}

\section{Comparison of critical temperatures to single-chain transition markers.}\label{SI_sec:markers}

For a solution at the critical temperature 
$T^*$ and critical volume fraction $\varphi ^*$, a slight perturbation in temperature will either result in a phase-separated or a homogeneous fluid. Given the prospective connection between the macroscopic phase transitions and single-chain physics, we compared temperatures associated with single-chains with critical temperatures. 
For the single-chain transitions, we performed simulations over a range of temperatures and extracted to estimates for transition temperatures. 
One is the temperature for which  $\tilde{\nu}\approx0.5$ (determined by interpolation); we denote this as 
$T^{\tilde{\theta}}$ by analogy to a theta-temperature.
The other temperatures considered coincide with maxima in heat capacity or energy fluctuations of the system; these are denoted as $T^{\text{max}}$.
The comparison for a particular systems associated with {\Rone}, {\Rtwo}, and {\Rthree} is shown in Table {\ref{tab_SI:2}}. 
Although we observe qualitative consistency in the number of transition temperatures, we do not otherwise identify a consistent relationship between the single-chain transition temperatures and the macroscopic phase-transition over this limited dataset.  

\begin{table}[H]
    \centering
        \caption{Comparison of computed critical temperature $T^*$ with single-chain transition temperatures, approximated by $T^{\tilde{\theta}}$ and $T^{\text{max}}$ for representatives systems with parameters from \Rone, \Rtwo, and \Rthree. }\label{tab_SI:2}
    \resizebox{0.6\textwidth}{!}{
    \begin{tabular}{|c|c|c|c|}
        \hline 
        Regime & Mean-field $T^*$ & $T^{\tilde{\theta}}$ & $T^{\text{max}}$ \\ 
        \hline 
        \Rone & 0.42 & 0.61 & 0.70 \\
        \Rtwo & 0.57, 16.95 & 0.54, 11.82 & 1.70, 10.0 \\
        \Rthree & 0.02, 0.15, 25.77 & 0.25, 0.35, 20.29 & 0.3, 1.0, 10\\
        \hline
    \end{tabular}}
\end{table}

\section{Parameters to reproduce experimental coexistence curves}
The versatility of the FHP framework extends to the construction of binodals with a myriad of shapes. Through the manipulation of its 8 intrinsic parameters, our framework achieves noteworthy concordance with experimental coexistence data, as illustrated in Figure 4. Employing the gradient-free covariance matrix adaptation strategy (CMA-ES), we have successfully identified parameter configurations that yield binodals congruent with empirical observations. The CMA-ES algorithm was implemented using the \textit{cma} module in python. 

The curves were obtained by solving
\begin{align}
    \Delta \mu _{\text{s}} &= \ln \left(\frac{1-\varphi _{\text{p}} ^\text{I}}{1-\varphi _{\text{p}} ^{\text{II}}}\right) + \left(1 - \frac{v_{\text{s}}}{v_{\text{p}}}\right)\left(\varphi _{\text{p}} ^{\text{I}} - \varphi _{\text{p}} ^{\text{II}} \right) + \chieff v_{\text{s}} \left(\left(\varphi _{\text{p}}^{\text{I} }\right) ^2 - \left(\varphi _{\text{p}}^{ \text{II} }\right) ^2\right)\\
    \Delta \mu _p &= \ln \left(\frac{\varphi _{\text{p}} ^\text{I}}{\varphi _{\text{p}} ^{\text{II}}}\right) + \left(1 - \frac{v_{\text{p}}}{v_{\text{s}}}\right)\left((1-\varphi _{\text{p}} ^{\text{I}}) - (1-\varphi _{\text{p}} ^{\text{II}}) \right) + \chieff v_{\text{p}} \left(\left(1-\varphi _{\text{p}}^{\text{I} }\right) ^2 - \left(1-\varphi _{\text{p}}^{ \text{II} }\right) ^2\right)
\end{align}

From \textit{cma} module the \textit{fmin} methods was used to execute the CMA-ES algorithm with the initial variance set to $0.01$, population size set to $10$, function tolerance \textit{tolfun} was set to $10^{-6}$, tolerance in $x$-values \textit{tolx} was set to $10^{-6}$ for $500$ iterations. 
\newpage 
\begin{sidewaystable}
    \centering
    \begin{table}[H]
        \caption{Parameters that reproduce the coexistence curves shown in Fig. 4. The experimental coexistence curves show a miscibiliy loop for a polyethylene glycol (PEG) in water (H$_2$O) and 2-butxoyethanol (2-BE) in water (H$_2$O); upper critical solution temperature (UCST) for polyisobutylene (PIB) with molecular weight 227000, 285000, 6,000,000 amu in diisobutyl ketone (DIBK), lower critical solution temperature (LCST) for polystyrene (PS) with molecular weight 100,000, 223,000 and 600,000 amu in ethylene acetate (EA), and an hourglass shape for polystyrene with molecular weight 100,000, 233,000, and 600,000 amu in tert-butyl acetate (TBA). }
        \label{tab:my_parameters}
        \resizebox{\columnwidth}{!}{
        \setlength{\tabcolsep}{0.1pt}
        \begin{tabular}{|c|c|c|c|c|c|c|c|c|c|c|c|}
            \hline 
            \multicolumn{1}{|c|}{} & \multicolumn{2}{|c|}{\textbf{Loop}} &  \multicolumn{3}{|c|}{\textbf{UCST}} & \multicolumn{3}{|c|}{\textbf{LCST}} & \multicolumn{3}{|c|}{\textbf{Hourglass}} \\ 
            \hline 
            \multicolumn{1}{|c|}{} & PEG/H$_2$O & 2-BE/H$_2$O &PIB(227k)/DIBK & PIB(285k)/DIBK & PIB(6m)/DIBK& PS(100k)/EA& PS(223k)/EA& PS(600k)/EA& PS(100k)/TBA& PS(233k)/TBA& PS(600k)/TBA\\
            \hline
            $p_v$ & 0.999922 & 1 & 0.402094 & 0.431279664	& 0.493237252 & 1 & 1 & 1 & 1 & 1 & 1 \\
            $p_w$ & 0.367884 & 0.351223	& 0.410100493 & 0.655368804	& 0.498172777 & 0.353040266 & 0.348741959 & 0.349214352	& 0.353429939	& 0.355928 & 0.353057925 \\
            $Mv_m$ & 156.124071 & 500.016557	& 151.0908664 & 1000.01773 & 4324.94932	& 170.027238 & 199.981543 & 500.005536 & 75.0115804 & 250.013337	& 599.932109 \\
            $v_s$ & 6.1376739 & 1.02725865 & 1.25438629	& 1.08880596 & 0.997370072 & 1.10606396	& 0.987025457 & 0.987975343 & 0.998016369	& 0.984431386 & 1.00053909 \\
            $\Emsa/k_{\text{B}}$ & -720.210927	& -533.73979 & -0.42745284 & -0.25637422 & 0.018211311 & -799.9735 & -738.068921 & -725.204044 & -2439.5288 & -2439.44825 & -2422.48224 \\ 
            $\Emsn/k_{\text{B}}$ & -141.873416	& -101.796721 & -1.15496282	& -0.205819368 & 0.028715109 & -131.521966 & -131.595669 & -131.552467 & -240.026102 & -239.966901 & -239.973644 \\
            $\Emma/k_{\text{B}}$ & -1167.24828 & -922.889237 & -13.7680371 & -14.2357021 & -14.1778536 & -1289.99258 & -1182.46617 & -1159.99088	& -4229.98574 & -4233.04005 & -4199.99049 \\ 
            $\Emmn /k_{\text{B}}$ & -130.93181 & -229.992313 & -13.31637158	& -13.3154217 & -14.2030666	& -169.924631 & -169.990196	& -169.941249	& -599.975512 & -600.006589	& -600.01121 \\
            $\Essa/k_{\text{B}}$ &  -118.30791	& -0.115343225 & -0.40768503 & -0.24642711 & -0.005601559 & -144.983301	& -145.007093 & -145.016402	& -768.011187 & -767.972788	& -768.024496\\
            $\Essn/k_{\text{B}}$ & -13.8431102	& 0.092407127 & -0.22232703	& 0.155786444 & 0.039231938	& -50.0038813 & -49.9775718	& -49.9835477	& -480.033377 & -480.047341	& -480.005756\\
            \hline 
        \end{tabular}}
    \end{table}
\end{sidewaystable}